\documentclass[12pt]{article}
\usepackage{amssymb}
\usepackage{natbib}
\usepackage{pdfwidgets}
\usepackage{cite}
\usepackage{longtable,lscape}
\usepackage{threeparttable}
\usepackage{tablefootnote}
\usepackage{verbatim}
\usepackage{graphicx}
\usepackage{geometry}
\usepackage{titling}
\pdfoutput=1
\bibpunct{(}{)}{;}{a}{}{,}
\begin{document}
\date{}
\setlength{\droptitle}{-5em}

%\begin{frontmatter}
\title{New methodology to determine the terminal height of a fireball\\
\bigskip
\normalsize
Manuel Moreno-Ib\'a\~nez$^{a,*}$, Maria Gritsevich$^{b,c,d,e}$, Josep M. Trigo-Rodr\'iguez$^a$\\
\bigskip
\small
\textit{$^a$Institute of Space Sciences (CSIC-IEEC), Campus UAB, Facultat de Ci\`encies, Torre C5-parell-$2^{nd}$ floor, 08193 Bellaterra, Barcelona, Spain}\\
\textit{$^b$Finnish Geospatial Research Institute (FGI), Department of Geodesy and Geodynamics, Geodeetinrinne 2, FI-02431 Masala, Finland }\\
\textit{$^c$Ural Federal University, Institute of Physics and Technology, Department of Physical Methods and Devices for Quality Control, Mira street 19, 620002 Ekaterinburg, Russia }\\
\textit{$^d$Russian Academy of Sciences, Dorodnicyn Computing Centre, Department of Computational Physics, Vavilova street 40, 119333 Moscow, Russia }\\
\textit{$^e$Moscow State University of Geodesy and Cartography (MIIGAiK), Extraterrestrial Laboratory (MExLab), Gorokhovsky pereulok 4, 105064 Moscow, Russia }\\
\textit{$^*$Corresponding author. E-mail address: mmoreno@csic.ice.es}\\
\bigskip
\bigskip
\normalsize
Accepted for publication in Icarus on December 21$^{st}$, 2014.}

\maketitle

%\author[ICE]{Manuel Moreno-Ib\'a\~nez} %\corref{cor1}} 
%%\email{mmoreno@csic.ice.es}

%\author[Helsinki,Moscow1,Moscow2,Moscow3]{Maria Gritsevich}
%\author[ICE]{Josep M. Trigo-Rodr\'iguez}

%\cortext[cor1]{Corresponding author. E-mail: mmoreno@ice.csic.es}
%\address[ICE]{Institute of Space Sciences (CSIC-IEEC), Campus UAB, Facultat de Ci\`encies, Torre C5-parell-$2^{nd}$ floor, 08193 Bellaterra, Barcelona, Spain}
%\address[Helsinki]{Finnish Geospatial Research Institute (FGI), Department of Geodesy and Geodynamics, Geodeetinrinne 2, FI-02431 Masala, Finland}
%\address[Moscow1]{Ural Federal University, Institute of Physics and Technology, Department of Physical Methods and Devices for Quality Control, Mira street 19, 620002 Ekaterinburg, Russia}
%\address[Moscow2]{Russian Academy of Sciences, Dorodnicyn Computing Centre, Department of Computational Physics, Vavilova street 40, 119333 Moscow, Russia}
%\address[Moscow3]{Moscow State University of Geodesy and Cartography (MIIGAiK), Extraterrestrial Laboratory (MExLab), Gorokhovsky pereulok 4, 105064 Moscow, Russia}
%\address[*]{Corresponding author. E-mail: mmoreno@ice.csic.es}

%\author[Insitution1]{Author1}
%\author[Insitution2]{Author2}

%\address[Insitution1]{Address1}
%\address[Insitution2]{Address2}

%\hline
\begin{abstract}

Despite ablation and drag processes associated with atmospheric entry of meteoroids were a subject of intensive study over the last century, little attention was devoted to interpret the observed fireball terminal height. This is a key parameter because  it not only depends on the initial mass, but also on the bulk physical properties of the meteoroids and hence on their ability to ablate in the atmosphere. In this work we have developed a new approach that is tested using the fireball terminal heights observed by the Meteorite Observation and Recovery Project operated in Canada between $1970-1985$ (hereafter referred as MORP). We then compare them to the calculation made. Our results clearly show that the new methodology is able to forecast the degree of deepening of meteoroids in the Earth's atmosphere. Then, this approach has important applications in predicting the impact hazard from cm- to meter-sized bodies that are represented, in part, in the MORP bolide list.

\end{abstract}
%\hline
%\begin{keyword}

%\textbf{Keywords:} Meteoroid \sep Meteors \sep Near-Earth objects \sep Impact processes



%\end{keyword}

%\end{frontmatter}


\section{Introduction}
\label{Intro}

\hspace{4ex}Deriving the meteoroid trajectories in the atmosphere is of particular interest to researchers. On the one hand, orbital parameters can be derived based on the time of appearance, meteor position and initial velocity, allowing us to estimate their parental relationship with parent bodies - asteroids and comets. On the other hand, a knowledge of physical parameters such as mass, velocity, deceleration, height, etc, at different points of its trajectory turns out to be very useful so as to predict the energy of a possible surface impact, locate meteorite fall and/or understand the ablation and other mass loss mechanisms occurring along the flight.

Various photographic and video techniques have been developed to obtain the most accurate and systematic observations of meteors. \citet{Whipple1957} (later modified by \citet{McCrosky1968}, \citet{Pecina1983} and \citet{Ceplecha1993}) derived a methodology by considering the separate meteor trails obtained when shuttering the video image. This technique allowed them to study the problem at shorter flight intervals. This analysis was dependent on the body properties' average values provided by the bibliography. However, a reliable theoretical flight model is still to be developed. The accuracy of theoretical results usually requires a very good precision in observation techniques. For example, \citet{Ceplecha1993} developed a theoretical model which included meteor fragmentation, but it needs very precise fireball records.

Along the years different methods have been developed in order to increase the accuracy of the theoretical models.

One of the first theoretical models for meteors (known as Single Body Theory)  was developed by \citet{Hoppe1937}. It was an extensive study of the flight mechanics and thermodynamics processes. \citet{Levin1956,Levin1961} studied the meteoroids' atmospheric entry with account for fragmentation and deceleration. He concluded that the mass of the body is related to its middle section by means of a parameter that characterizes the rotation of the fireball. 

Later on, \citet{Ceplecha1976} explored which fireballs of the Prairie Network were ordinary chondrites. The authors considered the fireball terminal height as a main characteristic factor. It was concluded that carbonaceous material shall ablate more readily and, consequently, these bodies may have shorter trajectories. The authors derived an empirical criterion (equation (\ref{eq0})) that established a weighted relation between the fireball terminal height and other flight properties of the fireball (namely air density at terminal height, $\rho_{E}$ $[g/cm^{3}]$, the preatmospheric mass, $m_{\infty}$ $[g]$, preatmospheric velocity, $V_{\infty}$ $[km/s]$, and the zenith distance of the meteor radiant, $Z_{R}$ $[degrees]$).

\begin{equation}
	PE = log\:\rho_{E}+A\;log\:m_{\infty}+B\;log\:V_{\infty}+C\;log\:(cos\:Z_{R})  \label{eq0}
\end{equation}

This expression also has a theoretical meaning based on the single body theory, as explained in their work. Coefficients A, B and C are obtained by using a least squares fit to $156$ fireballs of the Prairie Network \citep{McCrosky1965}. Owing to this criterion (equation (\ref{eq0})), \citet{Ceplecha1976} classified the Prairie Network fireballs into four different groups. In their discussion they suggested that ordinary chondrites should all belong to the same range of PE values ($-4.60<PE$). 

Besides, \citet{Ceplecha1976} took advantage of the fireballs' observed properties in order to shed more light in the validity of the previous criterion (equation (\ref{eq0})). Two parameters were used: $K$ (the shape-density coefficient) $[cm^{2}g^{-2/3}]$and $\sigma$ (ablation coefficient) $[s^{2}cm^{-2}]$. The average values (for all the observational measurements) are used to define a new parameter, SD:
\begin{equation}
	SD =\; <log\:K>\;+\;<log\:\sigma>  \label{eq01}
\end{equation}

The average numbers, $<log\:K>$ and $<log\:\sigma>$, are weighted by the ratio of the deceleration to its formal rms error. Then, SD is a parameter that has little influence from observational errors. As \citet{Ceplecha1976} stated, SD depends on the second derivative of the observational measurements via these two refined numbers, whereas PE is chosen as the simplest possible empirical expression. Therefore, SD is a new criterion which can be compared against the PE criterion.
These two parameters (\ref{eq0}) and (\ref{eq01}) turned out to be related when the meteor initial mass could be considered small or the ablation was large.

Slightly different methodology was suggested by \citet{Wetherill1981}. They included four meteorite selection criteria to build up their classification. \citet{Wetherill1981}, gave more importance to the dynamic mass than to the photometric mass; besides, they also took into account the deceleration of the body and the light curves to identify the survived meteorites among the fireballs registered by the Prairie Network. The authors highlighted the importance of the observed terminal height concluding that for meteorite-producing fireballs its value should agree with the theoretical value, calculated using dynamic mass, as well as with that of Lost City fireball to within $1.5$ km, when scaled for mass, velocity, and entry angle in accordance with classical single body meteor theory.

In \citet{Revelle1979}, the study of the interaction between large meteoroids and the atmosphere is done via a quasi-simple ablation model. The results are compared to photographically recorded meteorite falls as well. Later publications have gathered and expanded the physical problem of the deceleration of meteoroids in the atmosphere (e.g. \citet{Bronshten1983}).

\citet{ Halliday1989b, Halliday1989} studied observed fireball properties to derive the presence of correlations. They used 44 MORP recorded fireballs to classify as strong, moderate, weak or not having any correlation the observed data (i.e. initial velocity, total light emitted by the fireball, initial and end masses, initial and end heights, orbital elements, etc.). Despite of the observational errors (cameras not able to film all the trajectory, not clear sky, etc.), they found some strong correlations: mass lost by ablation versus the peak brightness, and  duration of luminosity recorded by MORP versus zenith distance of the radiant.

\citet{Stulov1995}, \citet{Stulov1997} and \citet{Gritsevich2007} proposed a new methodology. Instead of using the average values as input parameters, they gather all the unknown values into two new variables $\alpha$ (ballistic coefficient) and $\beta$ (mass loss parameter), mathematically introducing similar idea with scaling of parameters as suggested by \citet{Wetherill1981}. Adjusting the resulting equation to the trajectory observed, these new variables can be derived for each meteoroid. The resulting values allow to describe in details the meteoroid trajectory in the atmosphere and invent new classification scale for possible impacts \citep{Gritsevich2011b,Gritsevich2012}. This allows to determine other important parameters, such as preatmospheric and terminal mass values, ablation and shape change coefficients, as well as terminal height. The methodology to determine terminal height has been implemented for fully ablated fireballs by \citet{Gritsevich2008c}. In the present study we significantly specify and expand the applicability range of this methodology by testing it on larger data set.

In the following sections we present the results of applying this last methodology to a large number of MORP fireballs, including suspected meteorite-producing events included in the table 6 by \citet{Halliday1996}. Alternatively, we suggest a more accurate method of calculation. Section \ref{Theory} takes a look on previous and present terminal height determinations. We then compare observed values to our derived values in section \ref{Results}. Section \ref{Discussion} contains our discussion. Finally the conclusions and suggestions for future research are presented in section \ref{Conclusions}.

%---------------------------------------------------------------------------

\section{Theory}
\label{Theory}

\hspace{4ex}The equations of motion for a meteoroid entering the atmosphere projected onto the tangent and to the normal to the trajectory are well known. Once we consider some simplifications (see \citet{Gritsevich2010}) we have:

\begin{equation}
	M \frac{dV}{dt} = -\frac{1}{2}c_{d}\rho_{a}V^{2}S \label{eq4}
\end{equation}

\begin{equation} 
  \frac{dh}{dt} =-Vsin\gamma \label{eq5}
\end{equation}

$M$ is the body mass, $V$ is its velocity, $t$ is the time, $h$ is the height above the planetary surface, $\gamma$ is the local angle between the trajectory and the horizon, $S$ is the area of the middle section of the body, $\rho_{a}$ is the atmospheric density and $c_{d}$ is the drag coefficient.

Equations (\ref{eq4}), (\ref{eq5}) are complemented by the equation for the variable mass of the body:

\begin{equation}
	H^{*} \frac{dM}{dt} =-\frac{1}{2}c_{h}\rho_{a}V^{3}S, \label{eq6}
\end{equation}

where $H^{*}$ is the effective destruction enthalpy, and $c_{h}$ is the heat exchange coefficient. 
Two other expressions are added to solve the problem. On the one hand, we suggest that the atmosphere is isothermal, and so $\rho=exp(-h/h_{0})$, where $h_{0}$ is the scale height. On the other hand, from the research carried out by \citet{Levin1956,Levin1961} we can assume that the middle section and the mass of the body are connected by the following relation $s=m^{\mu}$, where $\mu$ is a constant. The parameter $\mu$ characterizes the possible role of rotation during the flight and can be calculated based on the observed brightness of a fireball using the method proposed by \citet{Gritsevich2011}. In \citet{Gritsevich2008} different values for $\mu$ are discussed. If $\mu=0$ there is no body rotation, whereas if $\mu=2/3$, the ablation of the body due to its rotation is uniform over the surface, and the shape factor does not change.  Generally we have $0<\mu<2/3$. According to the recent results \citep{Bouquet2014} the majority of the MORP fireballs were found to have $\mu$ values closer to $2/3$.

We now introduce dimensionless variables, $M=M_{e}m$, $V=V_{e}v$, $h=h_{0}y$, $\rho_{a}=\rho_{0}\rho$, $S=S_{e}s$. Here, the subscript $e$ indicates the parameters at the entry to the atmosphere, $h_{0}$ is a planetary scale height (we use $7.16 \cdot 10^{3}$ m for the Earth) and $\rho_{0}$ is the atmospheric density near the planetary surface. 

Taking into account all these previous considerations, we can turn the equation system (\ref{eq4})-(\ref{eq5})-(\ref{eq6}) into the following equations (see \citealp{Gritsevich2009}):

\begin{equation}
	m\frac{dv}{dy} = \alpha\rho vs \label{eq7}
\end{equation}

\begin{equation} 
  \frac{dm}{dy} = 2\alpha\beta \rho v^{2} s \label{eq8}
\end{equation}

These equations have been expressed as a function of two variables:

\begin{equation}
	\alpha=\frac{1}{2}c_{d}\frac{\rho_{0}h_{0}S_{e}}{M_{e}sin\gamma} \label{eq9}
\end{equation}

And

\begin{equation}
	\beta=(1-\mu)\frac{c_{h}V^{2}_{e}}{2c_{d}H^{*}}  \label{eq10}
\end{equation}

$\alpha$ is called the ballistic coefficient. It is proportional to the mass of the atmospheric column with the cross section $S_{e}$ along the trajectory divided by the body preatmospheric mass; it characterizes the drag intensity.

$\beta$ is known as the mass loss parameter.  It is proportional to the fraction of kinetic energy of the unit mass of the body that is transferred to the body in the form of heat divided by the effective destruction enthalpy.

The analytical solution of the equation system (\ref{eq7})-(\ref{eq8}) with the initial conditions $y=\infty$, $v=1$ and $m=1$ is:

\begin{equation}
	m= exp[-(1-v^{2})\beta /(1-\mu)] \label{eq11}
\end{equation}	

\begin{equation}
	y=ln2\alpha + \beta -ln\Delta,\  \  \Delta=\bar{Ei}(\beta)-\bar{Ei}(\beta v^{2}),  \label{eq12} 
\end{equation}

\begin{equation}
	\bar{Ei}(x)=\int^{x}_{-\infty} \frac{e^{t}dt}{t}  \label{eq12b}
\end{equation}

For quick meteors, one can neglect deceleration and set the velocity as constant for the luminous segment of the trajectory, then $v=V/V_{e}=1$. A strong evaporation process takes place so $\beta$ becomes high ($\beta>>1$). This means that the meteoroid ablation occurs rapidly in the absence of drag. For this situation, \citet{Stulov1998,Stulov2004} developed an asymptotic solution:

\begin{equation}
	v=1,\ \ m^{1-\mu}=1-2\alpha\beta e^{-y},\ \ ln2\alpha\beta<y<\infty \label{eq13}
\end{equation}

It should be noted that relation (\ref{eq13}) does not describe the drag process until the point where $m=0$. The meteor velocity begins to decrease in a certain vicinity of this point. In order to account for this change in velocity it was proposed to combine equation (\ref{eq13}) suitable for high $\beta$ values with equation (\ref{eq11}) suitable for arbitrary $\beta$ values (see
\citet{Gritsevich2008} and references therein). Such substitution of equation (\ref{eq11}) into equation (\ref{eq13}) fully escapes dependency of $\mu$ and gives the following height dependence of velocity at high $\beta$ values:

\begin{equation}
	v=\left( \frac{ln(1-2\alpha\beta e^{-y})}{\beta} +1        \right)^{1/2},\ \ ln2\alpha\beta<y<\infty \label{eq14}
\end{equation}

It is notable, that since condition of changing velocity was applied in the derivation of equation (\ref{eq14}), the equation (\ref{eq14}) is no longer applicable for constant-velocity meteors. Thus if condition $v=V/V_{e}=1$ is strictly set then solely equation (\ref{eq13}) should be considered.

In the opposite case, when $\beta=0$, the analytical solution (\ref{eq11})-(\ref{eq12}) becomes \citep{Gritsevich2008}:

\begin{eqnarray}
	m=1,\ \ y=ln\alpha-ln(-ln v) \label{eq15}
\end{eqnarray}

%%---------------------------------------------------------------------------------------

\subsection{Brief explanation of $\alpha$ and $\beta$ derivation}

\hspace{4ex}In their work, \citet{Gritsevich2007} explained the best way to derive $\alpha$ and $\beta$ parameters. Herein, we summarize the main points of the methodology. For a deeper explanation of the whole methodology we recommend a careful reviewing of the mentioned articles.

As explained in \citet{Gritsevich2008b} a weighted least-squared method is applied to the observed height and velocity values using equation (\ref{eq12}). Assigning manually the weighted factors may be quite complicated, so, since the height and velocity of a meteor decrease while it gets closer to the Earth's surface, the solution was proved to perform better if we take an exponential form of equation (\ref{eq12}). Mathematically it means that instead of direct comparison of $y_{i}$ (observed) with $y_{i}$ (calculated) the comparison is made between $e^{-y_{i}}$ values. The obtained results are then compared with the original data and a very good agreement is shown (see e.g. \citet{Gritsevich2008}). Then, equation (\ref{eq12}) turns into the following form:

\begin{eqnarray}
	2 \alpha exp(-y)-\Delta exp(-\beta)=0,\ \ \Delta=\bar{E}\bar{i}(\beta)-\bar{Ei}(\beta v^{2}), \\
	\bar{Ei}(x)=\int^{x}_{-\infty} \frac{e^{t}dt}{t}      \label{eq16}
\end{eqnarray}

$\alpha$ and $\beta$ are derived by minimizing the following expression (\ref{eq17}). We take into consideration that the analytical trajectory expression obtained shall accurately adjust the list of observed trajectory $(y_{i}, v_{i})$ values. 

\begin{equation}
	Q(\alpha,\beta)=\sum^{n}_{i=1}(F_{i}(y_{i},v_{i},\alpha,\beta))^{2}  \label{eq17}
\end{equation}

where

\begin{equation}
	F_{i}(y_{i}, v_{i}\alpha,\beta)= 2\alpha exp(-y_{i})-\Delta_{i} exp(-\beta) \label{eq18}
\end{equation}

In equation (\ref{eq17}) we assume $n\geq 3$, i.e. there are at least 3 points along the trajectory available for the calculation, one of which should contain initial velocity value, $V_{e}$, needed for scaling. It is also assumed that deceleration was noticeable during the luminous trail, i.e. there are at least 2 other than beginning points $(k, j \leq n)$ along the trajectory such as $V_{k}<V_{j}<V_{e}$. In principle, recent proliferation of instruments and methods dedicated to observing meteors makes it possible to fulfill this condition even for such small meteor particles as the Draconids \citep{Borovicka2007,Kero2012,Ye2013,Vaubaillon2015}.
In ordinary cases, when equally reliable data are provided along the observed trajectory (e.g. \citet{Halliday1996}), we suggest the use of all available $(h_{i}, V_{i})$ values, but this is not a strict requirement and specific initial filters on observational data may apply depending on studied case.

%---------------------------------------------------------------------------------------

\section{Results}
\label{Results}

\hspace{4ex}We have derived terminal heights by means of the previous development. In order to do so, we have used the $\alpha$ and $\beta$ values which have been previously derived using the methodology described above by \citet{Gritsevich2009} for the MORP fireballs. These heights are compared to the fireballs observed terminal height values; \citet{Halliday1996} collected them along with other fireball parameters such as: brightness, beginning and ending velocity, beginning height above sea, etc. 

The results of the following calculations are shown in table \ref{Table1}. The first row of this table contains the ID name of the fireball. Column $2$ contains $V_{e}$ which is the velocity of the body when it enters the atmosphere. Column $3$ shows the terminal velocity observed, $V_{t}$. Values for $V_{t}$ and $V_{e}$ are taken from \citet{Halliday1996}. The $\alpha$ and $\beta$ values appear in columns $4$ and $5$ respectively, and they are taken from \citet{Gritsevich2009}. The values of $h_{I}$, $h_{II}$ and $h_{III}$ in columns $6$, $7$ and $8$ correspond to the terminal heights derived in this study as we will see in the following lines. Finally, in column $9$ the observed terminal heights from \citet{Halliday1996} are listed for comparison with the values derived in our study.

\clearpage
%\onecolumn
%\begin{landscape}
\small
\begin{longtable}{l c c c c c c c c }
  \caption{Terminal Heights \label{Table1}}\\
 \hline \hline
   Fireball No. & $V_{e}$, km/s  & $V_{t}$, km/s & $\alpha$ & $\beta$ & $h_{I}$, km & $h_{II}$, km & $h_{III}$, km & $h_{obs}$, km \\ 
	\hline
\endfirsthead
\caption{Terminal Heights (continued)}\\ 
\hline\hline
  Fireball No.& $V_{e}$, km/s  & $V_{t}$, km/s & $\alpha$ & $\beta$ & $h_{I}$, km & $h_{II}$, km & $h_{III}$, km & $h_{obs}$, km \\ 
	\hline
\endhead
\hline
\multicolumn{9}{c}{Continued on next page}
\endfoot
\hline
\endlastfoot
   18 $^{[a]}$& 18.5 & 5.7 & 24.13 & 1.475 & 30.54 & 32.73 & 29.65 & 27.6 \\
    123 $^{[a]}$& 16.3 & 8.7 & 37.22 & 1.111 & 31.61 & 35.92 & 33.29 & 32.6 \\
    138 & 16.9 & 9.5 & 38.90 & 2.889 & 38.77 & 39.84 & 37.83 & 37.0 \\
    141 & 32.1 & 29.0 & 707.19 & 6.403 & 65.24 & 67.87 & 67.28 & 65.1 \\
    144 & 23.5 & 13.5 & 137.13 & 1.301 & 42.08 & 45.96 & 43.54 & 42.9 \\
    169 & 22.9 & 9.3 & 50.52 & 1.575 & 36.30 & 38.54 & 35.71 & 34.0 \\
    172 $^{[a]}$& 12.5 & 8.1 & 13.13 & 3.379 & 32.12 & 33.20 & 31.52 & 31.2 \\
    177 & 15.8 & 8.7 & 77.63 & 1.777 & 40.24 & 42.69 & 40.33 & 39.8 \\
    187 & 18.4 & 14.0 & 83.76 & 3.815 & 46.25 & 47.86 & 46.57 & 46.3 \\
    189 & 14.5 & 4.8 & 34.47 & 0.757 & 28.32 & 33.42 & 30.02 & 27.9 \\
    192 & 21.0 & 14.6 & 94.01 & 2.458 & 43.93 & 46.29 & 44.59 & 44.4 \\
    195 $^{[a]}$& 25.2 & 7.6 & 35.22 & 1.486 & 33.30 & 35.45 & 32.37 & 30.4 \\
    204 $^{[a]}$& 13.0 & 8.7 & 10.13 & 3.620 & 30.75 & 31.80 & 30.21 & 29.5 \\
    205 $^{[a]}$& 19.7 & 7.5 & 36.84 & 0.716 & 28.39 & 33.99 & 30.70 & 28.9 \\
    207 $^{[a]}$& 17.9 & 6.5 & 24.45 & 0.775 & 26.03 & 31.14 & 27.83 & 25.9 \\
    218 & 18.5 & 13.2 & 64.65 & 0.339 & 27.07 & 40.49 & 38.53 & 37.9 \\
    219 $^{[a]}$& 18.4 & 7.8 & 12.51 & 2.060 & 28.23 & 29.69 & 27.11 & 26.1 \\
    223 $^{[a]}$& 27.1 & 9.5 & 18.33 & 1.809 & 30.03 & 31.67 & 28.84 & 27.1 \\
    225 & 21.2 & 10.0 & 75.92 & 0.706 & 33.47 & 39.64 & 36.64 & 35.4 \\
    229 & 12.3 & 6.1 & 43.73 & 4.564 & 42.88 & 43.12 & 41.46 & 40.7 \\
    231 $^{[a]}$& 27.9 & 11.7 & 52.72 & 1.371 & 35.61 & 38.41 & 35.52 & 34.2 \\
    232 & 35.0 & 29.8 & 434.10 & 0 & \dots & \dots & 56.58 & 56.5 \\
    235 & 19.1 & 17.5 & 171.01 & 14.343 & 60.85 & 61.60 & 61.19 & 60.7 \\
    241 & 16.7 & 9.8 & 50.27 & 0.878 & 32.08 & 38.00 & 35.51 & 34.9 \\
    245 $^{[a]}$& 13.5 & 8.9 & 83.60 & 0.478 & 31.37 & 41.68 & 39.44 & 39.0 \\
    261 $^{[a]}$& 12.4 & 7.8 & 39.13 & 1.084 & 31.80 & 37.04 & 34.79 & 34.4 \\
    268 $^{[a]}$& 19.1 & 15.7 & 39.44 & 5.157 & 43.02 & 44.51 & 43.54 & 42.9 \\
    276 $^{[a]}$& 23.5 & 6.5 & 18.15 & 1.237 & 27.24 & 29.99 & 26.74 & 24.4 \\
    285 $^{[a,b]}$ & 14.54 & 2.7 & 8.25 & 1.700 & 23.87 & 25.41 & 22.30 & 19.8 \\
    287 & 23.4 & 10.5 & 40.85 & 1.750 & 35.53 & 37.57 & 34.91 & 34.6 \\
    288 $^{[a]}$& 12.4 & 4.1 & 9.05 & 1.215 & 22.13 & 25.09 & 21.93 & 20.2 \\
    299 $^{[a]}$& 23.6 & 14.7 & 48.36 & 1.042 & 33.03 & 38.41 & 36.12 & 35.7 \\
    300 & 21.5 & 10.2 & 202.36 & 2.071 & 48.20 & 49.80 & 47.33 & 46.2 \\
    301 & 36.1 & 24.4 & 693.65 & 0.662 & 48.85 & 57.42 & 55.30 & 55.6 \\
    303 $^{[a]}$& 14.1 & 8.2 & 44.15 & 1.260 & 33.74 & 37.82 & 35.41 & 34.9 \\
    304 & 16.2 & 6.6 & 48.17 & 2.442 & 39.10 & 40.10 & 37.64 & 36.8 \\
    307 $^{[a]}$& 21.0 & 3.8 & 12.08 & 1.760 & 26.85 & 28.29 & 25.21 & 22.0 \\
    310 & 17.0 & 10.5 & 67.74 & 0.559 & 30.98 & 39.79 & 37.36 & 36.8 \\
    313 $^{[a]}$& 16.7 & 10.2 & 134.20 & 0.432 & 34.03 & 44.33 & 41.83 & 41.1 \\
    314 & 14.6 & 9.3 & 159.93 & 0.118 & 26.00 & 45.27 & 42.83 & 42.1 \\
    331 & 13.3 & 7.0 & 37.94 & 0.598 & 27.32 & 34.81 & 31.98 & 31.1 \\
    340 & 29.1 & 25.6 & 1165.35 & 5.894 & 68.22 & 70.41 & 69.70 & 69.1 \\
    345 $^{[a]}$& 17.4 & 10.6 & 46.21 & 0.829 & 31.07 & 37.52 & 35.11 & 34.3 \\
    346 & 15.7 & 8.5 & 21.81 & 1.997 & 31.98 & 33.99 & 31.67 & 31.0 \\
    364 $^{[a]}$& 11.3 & 5.3 & 20.34 & 0.567 & 22.47 & 29.84 & 26.77 & 25.4 \\
    384 & 21.1 & 12.1 & 476.52 & 2.121 & 54.50 & 56.47 & 54.28 & 53.8 \\
    388 & 21.9 & 14.8 & 81.67 & 0.542 & 32.10 & 41.88 & 39.74 & 39.4 \\
    390 & 19.7 & 16.7 & 958.47 & 1.344 & 56.24 & 64.51 & 63.44 & 63.4 \\
    391 & 19.5 & 9.0 & 72.46 & 1.323 & 37.63 & 40.75 & 37.96 & 36.9 \\
    410 & 31.0 & 27.3 & 901.17 & 2.912 & 61.33 & 66.59 & 65.78 & 64.9 \\
    414 & 29.9 & 27.7 & 375.26 & 12.707 & 65.61 & 66.90 & 66.49 & 66.5 \\
    419 & 24.7 & 16.2 & 329.97 & 0.962 & 46.21 & 52.38 & 50.23 & 49.8 \\
    425 & 17.6 & 12.3 & 43.41 & 1.699 & 35.76 & 39.65 & 37.83 & 37.2 \\
    436 & 29.0 & 25.9 & 224.06 & 3.797 & 53.27 & 57.73 & 57.02 & 57.0 \\
    445 & 21.3 & 13.7 & 119.22 & 0.700 & 36.64 & 44.44 & 42.16 & 41.6 \\
    462 & 19.1 & 10.6 & 140.03 & 2.015 & 45.36 & 47.40 & 45.13 & 44.5 \\
    481 & 13.7 & 5.9 & 77.70 & 1.046 & 36.45 & 40.43 & 37.44 & 35.8 \\
    486 & 33.2 & 26.4 & 624.60 & 1.059 & 51.46 & 59.57 & 58.16 & 58.1 \\
    503 $^{[a]}$& 14.8 & 10.5 & 25.27 & 1.765 & 32.15 & 36.01 & 34.25 & 34.1 \\
    511 $^{[a]}$& 18.1 & 9.8 & 48.38 & 1.156 & 33.78 & 37.95 & 35.36 & 31.7 \\
    516 & 20.5 & 13.3 & 104.87 & 3.703 & 47.65 & 48.54 & 46.92 & 46.3 \\
    518 & 15.5 & 8.7 & 75.32 & 2.391 & 42.15 & 43.70 & 41.55 & 40.8 \\
    521 & 18.8 & 12.8 & 188.55 & 0.998 & 42.46 & 48.77 & 46.74 & 46.4 \\
    528 & 19.5 & 8.4 & 37.25 & 1.525 & 33.89 & 36.33 & 33.54 & 32.4 \\
    529 & 18.5 & 10.2 & 35.26 & 1.779 & 34.60 & 37.05 & 34.69 & 34.0 \\
    536 & 12.9 & 9.3 & 61.78 & 3.332 & 43.11 & 44.72 & 43.24 & 42.8 \\
    549 & 26.5 & 12.1 & 111.65 & 0.740 & 36.57 & 42.39 & 39.35 & 38.1 \\
    555 & 25.5 & 24.1 & 481.92 & 11.668 & 66.79 & 69.22 & 68.88 & 68.8 \\
    557 & 36.5 & 27.8 & 138.83 & 2.139 & 45.73 & 49.47 & 48.00 & 47.9 \\
    559 & 26.0 & 20.1 & 135.61 & 3.454 & 48.99 & 51.04 & 49.76 & 49.6 \\
    561 & 16.0 & 11.7 & 154.14 & 2.817 & 48.45 & 50.70 & 49.18 & 48.6 \\
    565 $^{[a]}$& 13.2 & 10.7 & 12.57 & 4.913 & 34.49 & 35.95 & 34.93 & 34.8 \\
    567 $^{[a]}$& 23.4 & 8.1 & 101.02 & 1.560 & 41.19 & 43.28 & 40.32 & 38.7 \\
    577 & 22.0 & 21.6 & 86.30 & 33.029 & 61.92 & 64.52 & 64.40 & 64.3 \\
    589 & 24.3 & 12.0 & 167.95 & 1.549 & 44.78 & 47.44 & 44.83 & 43.9 \\
    593 & 20.3 & 16.1 & 313.35 & 1.224 & 47.56 & 54.78 & 53.38 & 53.1 \\
    598 & 17.9 & 11.1 & 118.26 & 1.471 & 41.90 & 45.61 & 43.41 & 42.9 \\
    610 & 29.7 & 25.7 & 384.29 & 1.053 & 47.94 & 58.39 & 57.42 & 57.4 \\
    615 & 20.5 & 8.8 & 55.62 & 1.434 & 36.32 & 38.98 & 36.15 & 34.9 \\
    626 $^{[a]}$& 13.50 & 10.0 & 31.59 & 0.118 & 14.38 & 35.57 & 33.74 & 33.5 \\
    628 & 28.9 & 21.6 & 600.34 & 1.850 & 55.17 & 59.35 & 57.78 & 57.6 \\
    635 & 35.5 & 18.5 & 96.63 & 1.883 & 42.22 & 44.32 & 41.90 & 41.2 \\
    654 & 13.8 & 6.9 & 43.15 & 1.356 & 34.10 & 37.31 & 34.65 & 33.8 \\
    661 & 30.9 & 17.5 & 218.48 & 1.281 & 45.30 & 49.19 & 46.74 & 46.1 \\
    664 & 22.1 & 13.0 & 145.75 & 2.257 & 46.46 & 48.32 & 46.21 & 45.8 \\
    669 $^{[a]}$& 20.6 & 8.6 & 28.80 & 1.434 & 31.60 & 34.22 & 31.36 & 30.6 \\
    672 $^{[a]}$& 13.7 & 10.0 & 22.79 & 1.243 & 28.90 & 34.77 & 33.03 & 32.7 \\
    683 $^{[a]}$& 17.6 & 9.3 & 37.46 & 1.340 & 33.00 & 36.43 & 33.86 & 33.1 \\
    687 $^{[a]}$& 16.7 & 5.9 & 42.83 & 0.534 & 27.37 & 34.43 & 30.97 & 28.9 \\
    702 & 28.7 & 19.9 & 306.92 & 0 & \dots & \dots & 48.52 & 48.2 \\
    708 & 15.7 & 10.8 & 63.51 & 1.727 & 38.60 & 42.29 & 40.42 & 40.1 \\
    715 & 31.7 & 27.9 & 141.98 & 6.700 & 54.06 & 55.85 & 55.16 & 54.8 \\
    725 & 29.5 & 28.0 & 93.72 & 14.748 & 56.74 & 58.63 & 58.33 & 57.9 \\
    727 & 31.3 & 27.7 & 1160.73 & 2.100 & 60.80 & 68.00 & 67.20 & 67.2 \\
    741 & 31.9 & 28.5 & 370.61 & 3.764 & 56.81 & 61.32 & 60.61 & 60.6 \\
    751 & 18.9 & 8.9 & 44.19 & 0.377 & 25.10 & 34.91 & 31.78 & 30.5 \\
    752 & 13.6 & 11.9 & 8.67 & 9.418 & 36.48 & 37.32 & 36.69 & 36.5 \\
    763 & 27.6 & 25.2 & 254.30 & 4.410 & 55.24 & 59.93 & 59.34 & 59.4 \\
    769 & 17.4 & 10.5 & 50.00 & 1.233 & 34.47 & 38.84 & 36.52 & 36.0 \\
    771 $^{[a]}$& 17.1 & 7.8 & 41.75 & 0.731 & 29.44 & 35.32 & 32.28 & 31.0 \\
    774 & 33.0 & 27.1 & 85.66 & 0.994 & 36.78 & 45.99 & 44.74 & 44.7 \\
    776 & 13.1 & 10.0 & 33.21 & 5.497 & 42.25 & 43.01 & 41.89 & 41.6 \\
    792 $^{[a]}$& 17.3 & 7.7 & 37.43 & 0.388 & 24.12 & 33.57 & 30.34 & 28.9 \\
    819 & 17.7 & 11.1 & 123.90 & 1.034 & 39.71 & 45.18 & 42.90 & 42.4 \\
    823 & 25.0 & 19.4 & 32.61 & 4.927 & 41.33 & 42.42 & 41.28 & 40.0 \\
    829 & 28.9 & 15.1 & 359.02 & 0.869 & 46.08 & 51.51 & 48.76 & 47.7 \\
    835 & 24.2 & 8.2 & 66.13 & 2.474 & 41.46 & 42.31 & 39.77 & 38.4 \\
    840 & 23.6 & 6.1 & 22.35 & 1.766 & 31.28 & 32.81 & 29.81 & 27.7 \\
    843 & 36.5 & 33.7 & 642.92 & 4.526 & 62.07 & 67.22 & 66.69 & 66.6 \\
    844 $^{[a]}$& 14.9 & 9.8 & 23.89 & 1.694 & 31.46 & 34.91 & 32.92 & 32.6 \\
    852 & 15.9 & 4.6 & 21.27 & 2.541 & 33.53 & 34.26 & 31.69 & 29.9 \\
    857 & 18.5 & 12.0 & 62.52 & 1.224 & 36.02 & 40.87 & 38.74 & 38.3 \\
    860 & 23.3 & 17.6 & 330.46 & 0.283 & 37.46 & 52.98 & 51.25 & 51.1 \\
    861 & 13.3 & 7.2 & 49.82 & 2.720 & 40.11 & 41.24 & 39.14 & 38.6 \\
    865 & 26.5 & 18.1 & 258.85 & 0.564 & 40.65 & 50.29 & 48.20 & 47.8 \\
    868 & 27.8 & 25.6 & 1234.31 & 9.416 & 71.99 & 73.94 & 73.47 & 73.4 \\
    871 & 23.9 & 20.2 & 134.32 & 0 & \dots & \dots & 47.87 & 47.8 \\
    872 $^{[a]}$& 14.8 & 5.8 & 12.34 & 0.631 & 19.66 & 25.98 & 22.68 & 20.8 \\
    873 & 16.5 & 14.8 & 927.96 & 5.170 & 65.65 & 68.89 & 68.24 & 68.0 \\
    878 & 15.5 & 14.1 & 58.70 & 1.512 & 37.08 & 47.62 & 46.96 & 46.9 \\
    883 & 26.1 & 8.8 & 95.53 & 2.115 & 42.97 & 44.16 & 41.45 & 40.1 \\
    884 $^{[a]}$& 12.4 & 8.3 & 26.37 & 2.569 & 35.15 & 37.13 & 35.35 & 34.9 \\
    886 & 13.8 & 7.4 & 55.66 & 0 & \dots & \dots & 33.18 & 32.0 \\
    888 $^{[a]}$& 25.5 & 9.0 & 31.85 & 1.171 & 30.87 & 34.06 & 30.92 & 29.2 \\
    892 & 28.9 & 6.2 & 32.77 & 1.514 & 32.92 & 34.84 & 31.65 & 28.9 \\
    901 & 32.6 & 27.3 & 636.32 & 0.422 & 45.01 & 60.28 & 59.10 & 59.0 \\
    902 & 29.6 & 22.4 & 447.06 & 2.547 & 55.35 & 58.29 & 56.85 & 56.7 \\
    920 & 16.7 & 12.7 & 318.85 & 4.052 & 56.26 & 57.69 & 56.42 & 56.3 \\
    921 & 28.1 & 22.4 & 1077.86 & 8.721 & 70.47 & 70.77 & 69.96 & 69.3 \\
    925 $^{[a]}$& 26.4 & 4.2 & 42.87 & 1.503 & 34.79 & 36.67 & 33.42 & 29.8 \\
    929 & 35.0 & 30.1 & 342.27 & 2.944 & 54.48 & 58.95 & 58.03 & 58.0 \\
    930 $^{[a]}$& 13.4 & 9.5 & 38.75 & 1.052 & 31.51 & 37.94 & 36.06 & 35.6 \\
    933 & 15.4 & 11.1 & 43.80 & 1.779 & 36.15 & 40.12 & 38.41 & 38.0 \\
    937 $^{[a]}$& 17.9 & 6.4 & 40.19 & 0.881 & 30.50 & 34.96 & 31.69 & 29.8 \\
    947 & 24.9 & 24.0 & 771.47 & 32.917 & 77.58 & 78.31 & 78.13 & 77.9 \\
    966 & 31.3 & 27.8 & 1384.16 & 0.604 & 53.14 & 68.34 & 67.51 & 67.5 \\
    967 & 30.0 & 15.0 & 239.38 & 0 & \dots & \dots & 43.09 & 41.9 \\
    977 $^{[a]}$& 23.3 & 9.4 & 30.90 & 1.065 & 29.98 & 33.76 & 30.69 & 29.4 \\
    984 & 36.2 & 26.1 & 128.24 & 0 & \dots & \dots & 43.00 & 42.7 \\
    995 & 28.9 & 26.0 & 240.98 & 4.700 & 55.31 & 59.07 & 58.42 & 58.8 \\
    996 & 26.9 & 14.1 & 59.56 & 2.133 & 39.65 & 41.36 & 39.04 & 38.4 \\
    1010 & 18.4 & 10.1 & 71.58 & 1.378 & 37.84 & 41.28 & 38.79 & 38.1 \\
    1011 & 37.6 & 36.6 & 546.56 & 24.155 & 72.90 & 75.26 & 75.10 & 75.0 \\
\end{longtable} 
$^{[a]}${Probable meteorite fall according to \citet{Halliday1996}.}
$^{[b]}${This calculation corresponds to the Innisfree meteorite and it is based on the data from \citet{Halliday1981}, see \citet{Gritsevich2008}.}
%\end{landscape}
\normalsize

We have followed a step by step methodology. In first instance, terminal height is derived as the height corresponding to
the trajectory point where $m=0$ based on equation (\ref{eq13}). The substitution of condition $m=0$ into equation (\ref{eq13}) leads to:

\begin{equation}
	h_{I}=h_{0}\cdot y_{t}= h_{0} \cdot ln(2\alpha\beta)  \label{eq19}
\end{equation}

where $h_{0}=7.16 \cdot 10^{3}$ m is the scale height, and $y_{t}$ is the scaled terminal height.
 
As it can be seen in table \ref{Table1}, differences between observational and analytical values are remarkable (the standard deviation is $\sigma = 4.11$ km) and we conclude that the method proposed by \citet{Gritsevich2008c} cannot be extended to be applicable for decelerated fireballs without further modification, though it is usable to derive parameters of small meteoroids with constant velocity \citep{Popelenskaya2010}. Figure \ref{GraphicI} shows graphically the obtained differences between the observed and derived terminal height values based on equation (\ref{eq19}). As a reference, we have also plotted the line for which $h_{I}=h_{obs}$ ($h_{obs}$ are the terminal heights observed taken from \citet{Halliday1996}).
We then took into consideration that $v$ shall vary up along the trajectory. Terminal heights were obtained from the equation (\ref{eq14}):

\begin{equation}
	h_{II}=h_{0} \cdot y_{t}= h_{0} \cdot ln\left( \frac{2\alpha\beta}{\left(1-e^{\beta(v_{t}^{2}-1)}\right)} \right) \label{eq20}
\end{equation}

Once again, results can be found in column $7$ in table \ref{Table1}. Notice that $v_{t}=V_{t}/V_{e}$ (where $V_{t}$ is the terminal velocity observed, and $V_{e}$ is the velocity of the body when it enters the atmosphere) is the scaled terminal velocity. 
As we can see, though differences between derived and observed terminal heights ($h_{II},h_{obs}$) have decreased (the results' standard deviation is $\sigma = 1.52$ km) there is still a wide spread in the results. However, if we represent terminal heights derived vs terminal heights observed (figure \ref{GraphicII}), a global linear tendency can be found. This indicates that this methodology seems to obtain a reasonable first-order approximation.

One possible explanation for the quantitative differences obtained in these results is the use of the approximated function 
(\ref{eq14}) established for high $\beta$ values instead of using more general equation (\ref{eq12}). For easy comparison with equation (\ref{eq12}), the equation (\ref{eq14}) can be rewritten as:

\begin{equation}
y=\ln\left(\frac{2\alpha\beta}{\left(1-e^{\beta\left(v^2-1\right)}\right)}\right)  \label{eq20a}
\end{equation}

We have investigated either any improvement to this approximation itself, i.e. any possibility of adjusting the equation (\ref{eq14}) to better match equation (\ref{eq12}) could also bring better alternative solution to the problem studied here. To answer this question, we have used results recently obtained in \citep{Gritsevich2015}. The authors have examined the possibility of improving approximated function (\ref{eq14}). The functions (\ref{eq12}) and (\ref{eq20a}) do not have singularities and they are monotonous on the interval $0<v<1$. The dependency on $\alpha$ is the same (additive as $ln(\alpha)$) for both, equations (\ref{eq12}) and (\ref{eq20a}). Under these conditions a shift along parameter $\beta$ could quantitatively improve approximation represented by equation (\ref{eq20a}). And, in particularly, based on further functional analysis, it was found that much better approximation for the function (\ref{eq12}) can be achieved with the use of the function:

\begin{equation}
y= \ln\left(\frac{2\alpha\left(\beta-1.1\right)}{\left(1-e^{\left(\beta-
1.1\right)\left(v^2-1\right)}\right)}\right)   \label{eq20b}
\end{equation}

On figure ({\ref{Study}) we have graphically demonstrated the quality of the approximation relied on equation (\ref{eq20b}). We compare functions $y(\beta,v) = y(\alpha, \beta, v)-ln(2\alpha)$, where $y$  is taken according to equations (\ref{eq12}), (\ref{eq20a}) and (\ref{eq20b}). The comparison made for different $\beta$ values confirms that approximation (\ref{eq20b}) has better accuracy than previously used equation (\ref{eq20a}). In order to improve the results of the present study, we derived a new expression to calculate the terminal height of a fireball based on the referred above approximation (Eq. (\ref{eq20b}):

\begin{equation}
	h_{III}=h_{0} \cdot y_{t}= h_{0} \cdot ln\left( \frac{2\alpha(\beta-1.1)}{\left(1-e^{(\beta-1.1)(v_{t}^{2}-1)}\right)} \right) \label{eq21}
\end{equation}

A first justification to take up this further step is supported by figure \ref{GraphicIV}; it represents the differences ($h_{I}-h_{obs}$) vs. $\beta$. As expected, equation (\ref{eq19}) seems to give better results for larger $\beta$ values. The same happens with equation (\ref{eq20}), though in this case this dependency to higher $\beta$ values is smaller (as can be noticed from its standard deviation $\sigma = 1.52$ km).

Nonetheless, this new modification turns out to provide higher accuracy to our results. Now, the standard deviation becomes $\sigma = 0.75$ km. We can notice it graphically in figure \ref{GraphicIII}. As we can see, this figure shows the total agreement between derived and observed terminal heights using this new formula. Besides, it even allows us to derive terminal heights in those few cases where $\beta$ values are null.

%----------------------------------------------------------------------------------------------

\begin{figure}
\begin{center}
\input{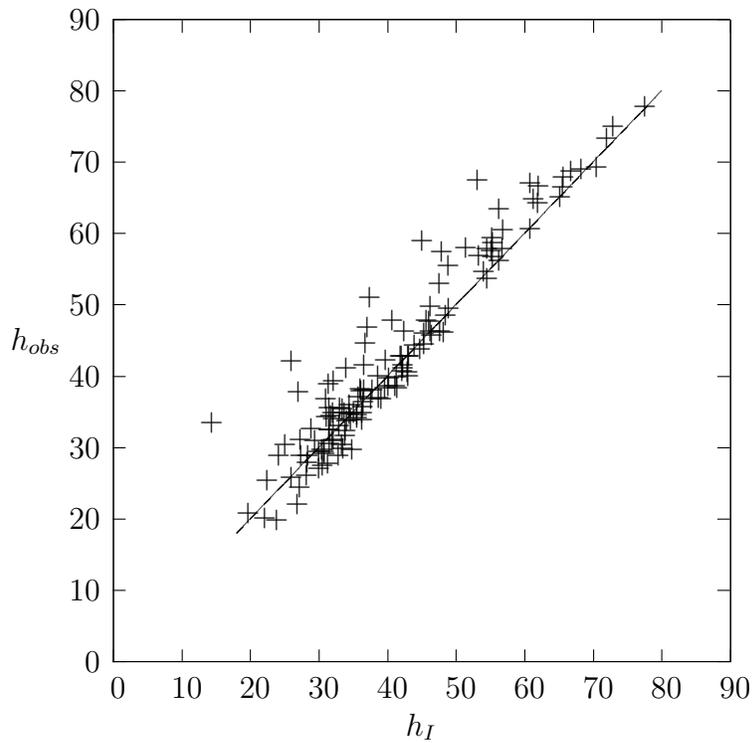}
\caption{Representation of derived terminal heights using equation (\ref{eq19}), $h_{I}$ versus the observed terminal height, $h_{obs}$. Line indicating $h_{I}=h_{obs}$ is also plotted.}
\label{GraphicI}
\end{center}
\end{figure}

\begin{figure}
\begin{center}
\input{graphicII.tex}
\caption{Representation of derived terminal heights using equation (\ref{eq20}), $h_{II}$ versus the observed terminal height, $h_{obs}$. Line indicating $h_{II}=h_{obs}$ is also plotted.}
\label{GraphicII}
\end{center}
\end{figure}

\begin{figure}
\begin{center}
\resizebox{\hsize}{!}{\includegraphics{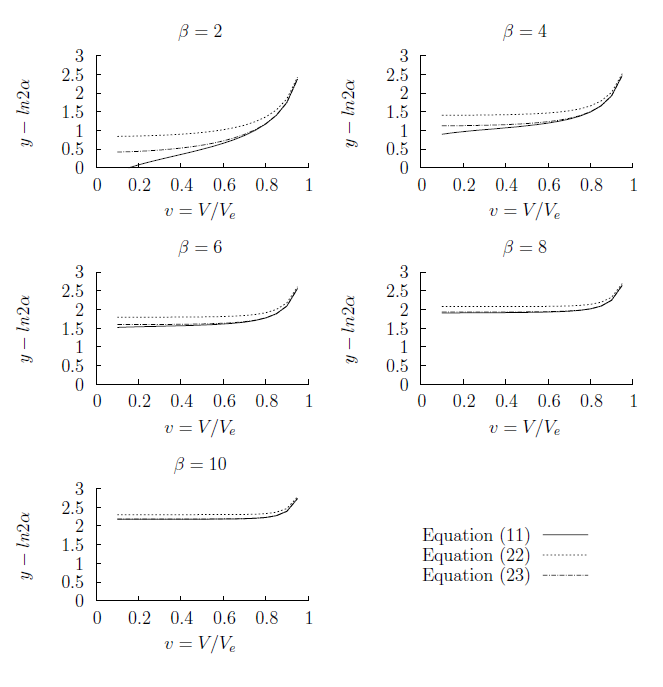}}
\caption{Graphical comparison of functions (\ref{eq12}), (\ref{eq20a}) and (\ref{eq20b}) for the selected $\beta$ values. The authors note, that for real fireballs, terminal velocity $v_{t}=V_{t}/V_{e}$ is usually higher than the one shown on the graphs (see experimental $V_{t}$ and $V_{e}$ values given in table \ref{Table1}), so only range $v_{t}<v<1$ would be applicable for practical purposes.}
\label{Study}
\end{center}
\end{figure}

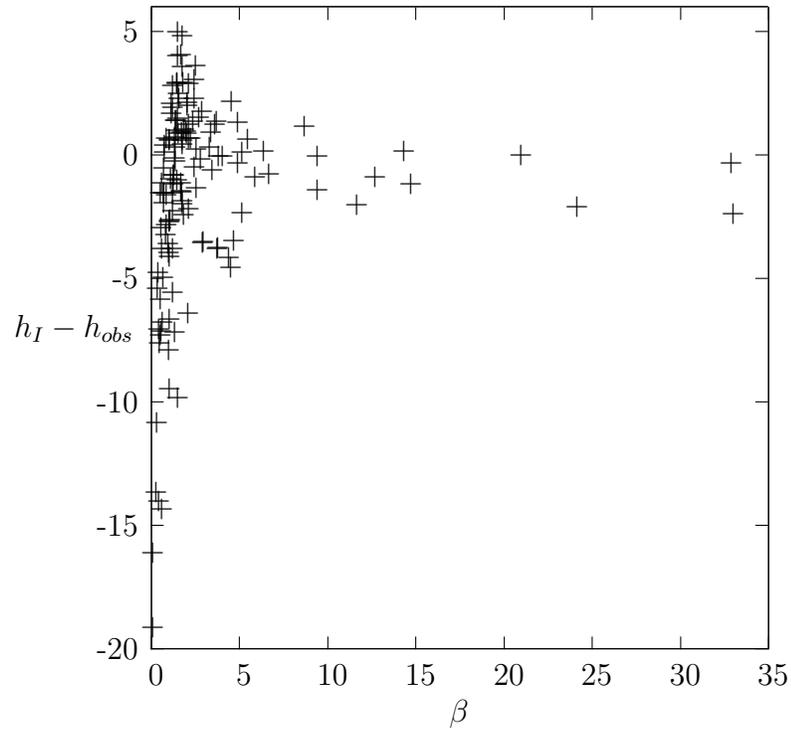
\begin{figure}
\begin{center}
% GNUPLOT: LaTeX picture
\setlength{\unitlength}{0.240900pt}
\ifx\plotpoint\undefined\newsavebox{\plotpoint}\fi
\sbox{\plotpoint}{\rule[-0.200pt]{0.400pt}{0.400pt}}%
\begin{picture}(1181,1181)(0,0)
\sbox{\plotpoint}{\rule[-0.200pt]{0.400pt}{0.400pt}}%
\put(151.0,131.0){\rule[-0.200pt]{4.818pt}{0.400pt}}
\put(131,131){\makebox(0,0)[r]{-20}}
\put(1100.0,131.0){\rule[-0.200pt]{4.818pt}{0.400pt}}
\put(151.0,325.0){\rule[-0.200pt]{4.818pt}{0.400pt}}
\put(131,325){\makebox(0,0)[r]{-15}}
\put(1100.0,325.0){\rule[-0.200pt]{4.818pt}{0.400pt}}
\put(151.0,519.0){\rule[-0.200pt]{4.818pt}{0.400pt}}
\put(131,519){\makebox(0,0)[r]{-10}}
\put(1100.0,519.0){\rule[-0.200pt]{4.818pt}{0.400pt}}
\put(151.0,713.0){\rule[-0.200pt]{4.818pt}{0.400pt}}
\put(131,713){\makebox(0,0)[r]{-5}}
\put(1100.0,713.0){\rule[-0.200pt]{4.818pt}{0.400pt}}
\put(151.0,907.0){\rule[-0.200pt]{4.818pt}{0.400pt}}
\put(131,907){\makebox(0,0)[r]{ 0}}
\put(1100.0,907.0){\rule[-0.200pt]{4.818pt}{0.400pt}}
\put(151.0,1101.0){\rule[-0.200pt]{4.818pt}{0.400pt}}
\put(131,1101){\makebox(0,0)[r]{ 5}}
\put(1100.0,1101.0){\rule[-0.200pt]{4.818pt}{0.400pt}}
\put(151.0,131.0){\rule[-0.200pt]{0.400pt}{4.818pt}}
\put(151,90){\makebox(0,0){ 0}}
\put(151.0,1120.0){\rule[-0.200pt]{0.400pt}{4.818pt}}
\put(289.0,131.0){\rule[-0.200pt]{0.400pt}{4.818pt}}
\put(289,90){\makebox(0,0){ 5}}
\put(289.0,1120.0){\rule[-0.200pt]{0.400pt}{4.818pt}}
\put(428.0,131.0){\rule[-0.200pt]{0.400pt}{4.818pt}}
\put(428,90){\makebox(0,0){ 10}}
\put(428.0,1120.0){\rule[-0.200pt]{0.400pt}{4.818pt}}
\put(566.0,131.0){\rule[-0.200pt]{0.400pt}{4.818pt}}
\put(566,90){\makebox(0,0){ 15}}
\put(566.0,1120.0){\rule[-0.200pt]{0.400pt}{4.818pt}}
\put(705.0,131.0){\rule[-0.200pt]{0.400pt}{4.818pt}}
\put(705,90){\makebox(0,0){ 20}}
\put(705.0,1120.0){\rule[-0.200pt]{0.400pt}{4.818pt}}
\put(843.0,131.0){\rule[-0.200pt]{0.400pt}{4.818pt}}
\put(843,90){\makebox(0,0){ 25}}
\put(843.0,1120.0){\rule[-0.200pt]{0.400pt}{4.818pt}}
\put(982.0,131.0){\rule[-0.200pt]{0.400pt}{4.818pt}}
\put(982,90){\makebox(0,0){ 30}}
\put(982.0,1120.0){\rule[-0.200pt]{0.400pt}{4.818pt}}
\put(1120.0,131.0){\rule[-0.200pt]{0.400pt}{4.818pt}}
\put(1120,90){\makebox(0,0){ 35}}
\put(1120.0,1120.0){\rule[-0.200pt]{0.400pt}{4.818pt}}
\put(151.0,131.0){\rule[-0.200pt]{0.400pt}{243.068pt}}
\put(151.0,131.0){\rule[-0.200pt]{233.432pt}{0.400pt}}
\put(1120.0,131.0){\rule[-0.200pt]{0.400pt}{243.068pt}}
\put(151.0,1140.0){\rule[-0.200pt]{233.432pt}{0.400pt}}
\put(30,635){\makebox(0,0){$h_{I}-h_{obs}$}}
\put(635,29){\makebox(0,0){$\beta$}}
\put(192,1021){\makebox(0,0){$+$}}
\put(182,869){\makebox(0,0){$+$}}
\put(231,976){\makebox(0,0){$+$}}
\put(328,913){\makebox(0,0){$+$}}
\put(187,875){\makebox(0,0){$+$}}
\put(195,996){\makebox(0,0){$+$}}
\put(245,943){\makebox(0,0){$+$}}
\put(200,924){\makebox(0,0){$+$}}
\put(257,905){\makebox(0,0){$+$}}
\put(172,923){\makebox(0,0){$+$}}
\put(219,889){\makebox(0,0){$+$}}
\put(192,1020){\makebox(0,0){$+$}}
\put(251,956){\makebox(0,0){$+$}}
\put(171,887){\makebox(0,0){$+$}}
\put(172,912){\makebox(0,0){$+$}}
\put(160,487){\makebox(0,0){$+$}}
\put(208,990){\makebox(0,0){$+$}}
\put(201,1021){\makebox(0,0){$+$}}
\put(171,832){\makebox(0,0){$+$}}
\put(277,992){\makebox(0,0){$+$}}
\put(189,962){\makebox(0,0){$+$}}
\put(732,907){\makebox(0,0){$+$}}
\put(548,913){\makebox(0,0){$+$}}
\put(175,798){\makebox(0,0){$+$}}
\put(164,611){\makebox(0,0){$+$}}
\put(181,806){\makebox(0,0){$+$}}
\put(294,912){\makebox(0,0){$+$}}
\put(185,1017){\makebox(0,0){$+$}}
\put(198,1065){\makebox(0,0){$+$}}
\put(199,943){\makebox(0,0){$+$}}
\put(185,982){\makebox(0,0){$+$}}
\put(180,804){\makebox(0,0){$+$}}
\put(208,985){\makebox(0,0){$+$}}
\put(169,645){\makebox(0,0){$+$}}
\put(186,862){\makebox(0,0){$+$}}
\put(219,996){\makebox(0,0){$+$}}
\put(200,1095){\makebox(0,0){$+$}}
\put(166,681){\makebox(0,0){$+$}}
\put(163,633){\makebox(0,0){$+$}}
\put(154,282){\makebox(0,0){$+$}}
\put(168,760){\makebox(0,0){$+$}}
\put(314,873){\makebox(0,0){$+$}}
\put(174,782){\makebox(0,0){$+$}}
\put(206,945){\makebox(0,0){$+$}}
\put(167,793){\makebox(0,0){$+$}}
\put(210,934){\makebox(0,0){$+$}}
\put(166,624){\makebox(0,0){$+$}}
\put(188,629){\makebox(0,0){$+$}}
\put(188,935){\makebox(0,0){$+$}}
\put(232,769){\makebox(0,0){$+$}}
\put(503,873){\makebox(0,0){$+$}}
\put(178,768){\makebox(0,0){$+$}}
\put(198,851){\makebox(0,0){$+$}}
\put(256,762){\makebox(0,0){$+$}}
\put(170,715){\makebox(0,0){$+$}}
\put(207,941){\makebox(0,0){$+$}}
\put(180,932){\makebox(0,0){$+$}}
\put(180,649){\makebox(0,0){$+$}}
\put(200,831){\makebox(0,0){$+$}}
\put(183,988){\makebox(0,0){$+$}}
\put(254,960){\makebox(0,0){$+$}}
\put(217,960){\makebox(0,0){$+$}}
\put(179,754){\makebox(0,0){$+$}}
\put(193,965){\makebox(0,0){$+$}}
\put(200,930){\makebox(0,0){$+$}}
\put(243,919){\makebox(0,0){$+$}}
\put(171,848){\makebox(0,0){$+$}}
\put(474,829){\makebox(0,0){$+$}}
\put(210,823){\makebox(0,0){$+$}}
\put(247,883){\makebox(0,0){$+$}}
\put(229,901){\makebox(0,0){$+$}}
\put(287,895){\makebox(0,0){$+$}}
\put(194,1004){\makebox(0,0){$+$}}
\put(1065,815){\makebox(0,0){$+$}}
\put(194,941){\makebox(0,0){$+$}}
\put(185,692){\makebox(0,0){$+$}}
\put(192,868){\makebox(0,0){$+$}}
\put(180,540){\makebox(0,0){$+$}}
\put(191,962){\makebox(0,0){$+$}}
\put(154,165){\makebox(0,0){$+$}}
\put(202,813){\makebox(0,0){$+$}}
\put(203,947){\makebox(0,0){$+$}}
\put(189,919){\makebox(0,0){$+$}}
\put(186,876){\makebox(0,0){$+$}}
\put(213,933){\makebox(0,0){$+$}}
\put(191,946){\makebox(0,0){$+$}}
\put(185,760){\makebox(0,0){$+$}}
\put(188,903){\makebox(0,0){$+$}}
\put(166,848){\makebox(0,0){$+$}}
\put(199,849){\makebox(0,0){$+$}}
\put(336,878){\makebox(0,0){$+$}}
\put(559,862){\makebox(0,0){$+$}}
\put(209,659){\makebox(0,0){$+$}}
\put(255,760){\makebox(0,0){$+$}}
\put(161,698){\makebox(0,0){$+$}}
\put(412,906){\makebox(0,0){$+$}}
\put(273,746){\makebox(0,0){$+$}}
\put(185,848){\makebox(0,0){$+$}}
\put(171,847){\makebox(0,0){$+$}}
\put(179,600){\makebox(0,0){$+$}}
\put(303,932){\makebox(0,0){$+$}}
\put(162,722){\makebox(0,0){$+$}}
\put(180,803){\makebox(0,0){$+$}}
\put(287,959){\makebox(0,0){$+$}}
\put(175,844){\makebox(0,0){$+$}}
\put(219,1026){\makebox(0,0){$+$}}
\put(200,1046){\makebox(0,0){$+$}}
\put(276,731){\makebox(0,0){$+$}}
\put(198,863){\makebox(0,0){$+$}}
\put(221,1048){\makebox(0,0){$+$}}
\put(185,819){\makebox(0,0){$+$}}
\put(159,378){\makebox(0,0){$+$}}
\put(226,966){\makebox(0,0){$+$}}
\put(167,630){\makebox(0,0){$+$}}
\put(412,852){\makebox(0,0){$+$}}
\put(168,863){\makebox(0,0){$+$}}
\put(294,816){\makebox(0,0){$+$}}
\put(193,526){\makebox(0,0){$+$}}
\put(210,1019){\makebox(0,0){$+$}}
\put(222,917){\makebox(0,0){$+$}}
\put(183,972){\makebox(0,0){$+$}}
\put(193,1063){\makebox(0,0){$+$}}
\put(163,364){\makebox(0,0){$+$}}
\put(222,855){\makebox(0,0){$+$}}
\put(263,906){\makebox(0,0){$+$}}
\put(392,953){\makebox(0,0){$+$}}
\put(193,1101){\makebox(0,0){$+$}}
\put(233,771){\makebox(0,0){$+$}}
\put(180,748){\makebox(0,0){$+$}}
\put(200,835){\makebox(0,0){$+$}}
\put(175,934){\makebox(0,0){$+$}}
\put(1062,895){\makebox(0,0){$+$}}
\put(168,350){\makebox(0,0){$+$}}
\put(180,930){\makebox(0,0){$+$}}
\put(281,772){\makebox(0,0){$+$}}
\put(210,956){\makebox(0,0){$+$}}
\put(189,897){\makebox(0,0){$+$}}
\put(820,826){\makebox(0,0){$+$}}
\put(151.0,131.0){\rule[-0.200pt]{0.400pt}{243.068pt}}
\put(151.0,131.0){\rule[-0.200pt]{233.432pt}{0.400pt}}
\put(1120.0,131.0){\rule[-0.200pt]{0.400pt}{243.068pt}}
\put(151.0,1140.0){\rule[-0.200pt]{233.432pt}{0.400pt}}
\end{picture}
\caption{Representation of the differences between derived and observed terminal heights ($h_{I}-h_{obs}$) versus $\beta$ values.}
\label{GraphicIV}
\end{center}
\end{figure}

\begin{figure}
\begin{center}
\input{graphicIII.tex}
\caption{Representation of derived terminal heights using equation (\ref{eq21}), $h_{III}$ versus the observed terminal height, $h_{obs}$. Line indicating $h_{III}=h_{obs}$ is also plotted.}
\label{GraphicIII}
\end{center}
\end{figure}

\section{Discussion}
\label{Discussion}

\hspace{4ex}As explained in section \ref{Theory}, equation (\ref{eq19}) should lead to good results when $\beta$ values are high. In figure \ref{GraphicIV} we notice that for $\beta > 5$ the differences between derived and observed terminal heights are small. However, this equation did not take into account the decrease in velocity close to the terminal point of the trajectory. Given that the amount of fireballs studied is large, we continued our analysis considering this fact (see equation \ref{eq20}).

Formula (\ref{eq20}) seemed to provide accurate results with regard to the observations; in fact, a clear global lineal tendency was found on the graph with the observed and calculated values of terminal height. Nonetheless, remarkable discrepancies are present along the whole range of $\beta$ values. No exceptions for higher, lower or intermediate $\beta$ values (see table \ref{Table1}).

The modification made by introducing new recently proposed approximation \citep{Gritsevich2015} in our methodology leads to higher accuracy.
This result may create some astonishment. Given that observations could carry different kind of errors: bad astrometric measurements, incomplete recording of the luminous trajectory of the meteor, blurring effects, etc., it should be expected that some of the analytical values differed from the observed ones. This could be explained if we remind that the differences obtained are of the order of $1$ km. Most of the aforementioned observational errors shall introduce smaller distance errors so these effects shall be comprised in our calculations.

Additional observation can be made based on figure (\ref{GraphicIII}) that equation (\ref{eq21}) produces its best fit for $h_{III}$ for the fireballs which terminated at higher altitudes. This result is expected since fireballs with survived terminal mass penetrate lower in the atmosphere, and they have, on average, lower values of mass loss parameter $\beta$. For example, \citet{Halliday1989b} derived dynamic mass estimation only late in the trail for those MORP fireballs that penetrated below a height of 40 km. \citet{Gritsevich2008d} calculated dynamic terminal masses of MORP fireballs based on the whole observed trajectories regardless of the terminal height value; indeed, all larger objects and/or suspected meteorite droppers penetrated below 40 km, with an average terminal height of 29 km. Thus, for planetary defense applications one can note, that larger objects could initiate fireballs penetrating deeper in the atmosphere than our methodology would predict.

It is also notable, that meteor height may be expressed as a function of time. Thus, the ability of predicting a terminal height may be directly linked with the forecast of a total duration of meteor phase. This can create scope to a new class of problems, such as, for example, insights into determination of luminous efficiency based on meteor duration and calculation of critical kinetic energy needed to produce luminosity. In future, several other tightly related problems may be considered, including a possibility of initial meteor height determination and deriving parameters $\alpha$ and $\beta$ without the beginning part of a meteor trail. Furthermore, the existing solution to the inverse problem of deriving these parameters described in section \ref{Theory} of this study is only applicable for the decelerated fireballs. There are still large amount of archived data and many new meteor registrations for which the recorded deceleration was not significant and/or it is in the order of observational errors. For these cases observed terminal heights will be crucial in setting margins on possible $\alpha$ and $\beta$ values based on equation (\ref{eq19}). Some initial steps in solving this problem were taken in \citet{Popelenskaya2010}.

Moreover, even for obviously decelerated and well-studied cases, such as Innisfree meteorite (fireball No. 285 in table \ref{Table1}), published terminal heights may differ depending on a taken data-reduction approach (e.g. 21 km in \citet{Halliday1981}; 19.8 km in \citet{Halliday1996}) and, therefore, any additional theoretical constrains on the terminal height and fireball duration may be very helpful.

Finally, it is interesting to compare our results with PE criterion introduced by \citet{Ceplecha1976} reviewed in the Introduction section of our study. In our methodology we used a different approach which, in particular, avoids A, B, C coefficients needed to be empirically set in PE criterion. We operated with dimensionless parameters $\alpha$ and $\beta$ which by their definition (equations (\ref{eq9}) and (\ref{eq10})) gather all the unknowns. The terminal height was a key sought parameter without passing to air density calculated at terminal height. We adjusted a theoretical expression which can be used to calculate terminal height. To verify the proposed dependency of terminal height on parameters $\alpha$ and $\beta$ we calculated terminal heights for MORP fireballs with known (observed and reported by \citet{Halliday1996}) terminal height and known values of $\alpha$ and $\beta$ parameters. However, if one would substitute $\alpha$ and $\beta$ in our criteria with their defining expressions (equations (\ref{eq9}) and (\ref{eq10})), one could find many similarities with PE criterion. Indeed, $\alpha$ depends on the preatmospheric cross-section-to-mass ratio (the ratio which can be easy converted into dependency on bulk density, preatmospheric mass and shape coefficient) and trajectory slope $\gamma$ related to $Z_{R}$. Mass loss parameter is proportional to preatmospheric velocity in power of 2 and inversely proportional to the effective destruction enthalpy. Thus, surpassing unknowns with described in our study methodology we find similar set of physical parameters affecting the degree of deepening of meteoroids in the Earth's atmosphere.

%------------------------------------------------------------------------------

\section{Conclusions}
\label{Conclusions}

\hspace{4ex}Previous work has been done to estimate the terminal height of fireballs (see section \ref{Intro}). In this paper, we have derived the terminal heights for MORP fireballs using newly developed as well as previously suggested methodology. This methodology had only been tested on several fully ablated fireballs with large $\beta$ values by \citet{Gritsevich2008c}. We were particularly interested in determining whether this new mathematical approach works equally accurately with fully ablated fireballs and meteorite-producing ones, and whether any additional modification could be applied to improve the accuracy and extend applicability range for the proposed methodology.

The comparison of the derived terminal heights of MORP fireballs with their observed values brings us to the following conclusions. Using equation (\ref{eq20}) it comes out that the accuracy shows a standard deviation of $\sigma = 1.52$ km which seems to be high. However, a linear tendency is found. This last reason encouraged us to find a better adjustment. By applying newly suggested approximation in our study equation (\ref{eq21}) we obtain terminal heights that fit very accurately the observed values; now the standard deviation for MORP data set turns out to be $\sigma = 0.75$ km.

Besides, the excellent adjustment between derived and observed terminal height values justifies the methodology proposed here. We foresee a calculation of terminal height to be useful when the lower part of the trajectory was not instrumentally registered (e.g. limited field of view of the camera, shadowing of fireball trail by other objects, weather conditions, detection of meteoroids using radar techniques with limited observation beam etc.). It also brings critical knowledge into the problem when one needs to predict how long will be a total duration of the luminous flight or at which height a fireball produced by a meteoroid with given properties would terminate. Based on our investigations we can highly recommend the use of equation (\ref{eq21}) also to solve inverse problem when terminal height and velocity are available from the observations, and parameters $\alpha$ and $\beta$ need to be derived. 

Further observational and analytical research must be done to predict precisely the properties of any body along its atmospheric flight. 

%------------------------------------------------------

\section*{Acknowledgments}

\hspace{4ex}We thank support from the Spanish Ministry of Science and Innovation (project AYA2011-26522), the Academy of Finland (project 260027) and the Finnish Geodetic Institute. The concept of the methodology and calculations were made by Manuel Moreno-Ib\'a\~nez and Maria Gritsevich, the later while she was hosted in MIIGAiK as invited expert in the project No. 14-22-00197 "Studies of Fundamental Geodetic Parameters and Topography of Planets and Satellites" supported by the Russian Science Foundation. Josep M. Trigo-Rodr\'iguez supervised and granted the study from the Spanish side. All authors contributed to the interpretation of the results and writing the manuscript.

We are grateful to the reviewers for their comments and constructive suggestions.

%Maria Gritsevich is supported by the Academy of Finland, part of the work was carried out in MIIGAiK and supported by the Russian Science Foundation, project No. 14-22-00197.

%------------------------------------------------------

%\section*{References}



\bibliographystyle{elsarticle-harv} 
\bibliography{bibliography}

\begin{thebibliography}{37}
\expandafter\ifx\csname natexlab\endcsname\relax\def\natexlab#1{#1}\fi
\expandafter\ifx\csname url\endcsname\relax
  \def\url#1{\texttt{#1}}\fi
\expandafter\ifx\csname urlprefix\endcsname\relax\def\urlprefix{URL }\fi

\bibitem[{{Borovi{\v c}ka} et~al.(2007){Borovi{\v c}ka}, {Spurn{\'y}}, and
  {Koten}}]{Borovicka2007}
{Borovi{\v c}ka}, J., {Spurn{\'y}}, P., {Koten}, P., Oct. 2007. {Atmospheric
  deceleration and light curves of Draconid meteors and implications for the
  structure of cometary dust}. Astronomy and Astrophysics 473, 661--672.

\bibitem[{Bouquet et~al.(2014)Bouquet, Baratoux, Vaubaillon, Gritsevich,
  Mimoun, Mousis, and Bouley}]{Bouquet2014}
Bouquet, A., Baratoux, D., Vaubaillon, J., Gritsevich, M.~I., Mimoun, D.,
  Mousis, O., Bouley, S., 2014. Simulation of the capabilities of an orbiter
  for monitoring the entry of interplanetary matter into the terrestrial
  atmosphere. Planetary and Space Science 103, 238--249.

\bibitem[{{Bronshten}(1983)}]{Bronshten1983}
{Bronshten}, V.~A., 1983. {Physics of Meteoric Phenomena}. D. Reidel Publishing
  Company, Dordrecht.

\bibitem[{{Ceplecha} and {McCrosky}(1976)}]{Ceplecha1976}
{Ceplecha}, Z., {McCrosky}, R.~E., Dec. 1976. {Fireball end heights - A
  diagnostic for the structure of meteoric material}. Journal of Geophysical
  Research 81, 6257--6275.

\bibitem[{{Ceplecha} et~al.(1993){Ceplecha}, {Spurny}, {Borovicka}, and
  {Keclikova}}]{Ceplecha1993}
{Ceplecha}, Z., {Spurny}, P., {Borovicka}, J., {Keclikova}, J., Nov. 1993.
  {Atmospheric fragmentation of meteoriods}. Astronomy and Astrophysics 279,
  615--626.

\bibitem[{Gritsevich and Koschny(2011)}]{Gritsevich2011}
Gritsevich, M., Koschny, D., 2011. Constraining the luminous efficiency of
  meteors. Icarus 212~(2), 877 -- 884.

\bibitem[{Gritsevich et~al.(2012)Gritsevich, Stulov, and
  Turchak}]{Gritsevich2012}
Gritsevich, M., Stulov, V., Turchak, L., 2012. Consequences of collisions of
  natural cosmic bodies with the earth’s atmosphere and surface. Cosmic
  Research 50~(1), 56--64.

\bibitem[{{Gritsevich}(2007)}]{Gritsevich2007}
{Gritsevich}, M.~I., Dec. 2007. {Approximation of the Observed Motion of
  Bolides by the Analytical Solution of the Equations of Meteor Physics}. Solar
  System Research 41~(6), 509--514.

\bibitem[{{Gritsevich}(2008{\natexlab{a}})}]{Gritsevich2008d}
{Gritsevich}, M.~I., Nov. 2008{\natexlab{a}}. {Estimating the terminal mass of
  large meteoroids}. Doklady Physics 53, 588--594.

\bibitem[{{Gritsevich}(2008{\natexlab{b}})}]{Gritsevich2008b}
{Gritsevich}, M.~I., 2008{\natexlab{b}}. {Identification of Fireball Dynamic
  Parameters}. Moscow University Mechanics Bulletin 63~(1), 1--5.

\bibitem[{{Gritsevich}(2008{\natexlab{c}})}]{Gritsevich2008}
{Gritsevich}, M.~I., Oct. 2008{\natexlab{c}}. {The Pribram, Lost City,
  Innisfree, and Neuschwanstein falls: An analysis of the atmospheric
  trajectories}. Solar System Research 42, 372--390.

\bibitem[{{Gritsevich}(2009)}]{Gritsevich2009}
{Gritsevich}, M.~I., Aug. 2009. {Determination of parameters of meteor bodies
  based on flight observational data}. Advances in Space Research 44, 323--334.

\bibitem[{{Gritsevich}(2010)}]{Gritsevich2010}
{Gritsevich}, M.~I., Aug. 2010. {On a Formulation of Meteor Physics Problems}.
  Moscow University Mechanics Bulletin 65~(4), 94--95.

\bibitem[{{Gritsevich} et~al.(2015){Gritsevich}, {Lukashenko}, and
  {Turchak}}]{Gritsevich2015}
{Gritsevich}, M.~I., {Lukashenko}, V.~T., {Turchak}, L.~I., 2015.
  {Approksimatsiya resheniya uravneniy meteornoy fiziki elementarnymi
  funkstsiyami (Elementary functions approximation for the solution of meteor
  physics equations)}. Matematicheskoe Modelirovanie 27~(2), 25--33.

\bibitem[{{Gritsevich} and {Popelenskaya}(2008)}]{Gritsevich2008c}
{Gritsevich}, M.~I., {Popelenskaya}, N.~V., Feb. 2008. {Meteor and fireball
  trajectories for high values of the mass loss parameter}. Doklady Physics 53,
  88--92.

\bibitem[{{Gritsevich} et~al.(2011){Gritsevich}, {Stulov}, and
  {Turchak}}]{Gritsevich2011b}
{Gritsevich}, M.~I., {Stulov}, V.~P., {Turchak}, L.~I., Mar. 2011. {Standards
  for crater formation and meteorite fallout by the light sector of an
  atmospheric trajectory}. Doklady Physics 56, 199--203.

\bibitem[{{Halliday} et~al.(1989{\natexlab{a}}){Halliday}, {Blackwell}, and
  {Griffin}}]{Halliday1989b}
{Halliday}, I., {Blackwell}, A.~T., {Griffin}, A.~A., Apr. 1989{\natexlab{a}}.
  {Detailed records of many unrecovered meteorites in western Canada for which
  further searches are recommended}. Journal of the Royal Astronomical Society
  of Canada 83, 49--80.

\bibitem[{{Halliday} et~al.(1989{\natexlab{b}}){Halliday}, {Blackwell}, and
  {Griffin}}]{Halliday1989}
{Halliday}, I., {Blackwell}, A.~T., {Griffin}, A.~A., Jun. 1989{\natexlab{b}}.
  {The typical meteorite event, based on photographic records of 44 fireballs}.
  Meteoritics 24, 65--72.

\bibitem[{{Halliday} et~al.(1981){Halliday}, {Griffin}, and
  {Blackwell}}]{Halliday1981}
{Halliday}, I., {Griffin}, A.~A., {Blackwell}, A.~T., Jun. 1981. {The Innisfree
  meteorite fall - A photographic analysis of fragmentation, dynamics and
  luminosity}. Meteoritics 16, 153--170.

\bibitem[{{Halliday} et~al.(1996){Halliday}, {Griffin}, and
  {Blackwell}}]{Halliday1996}
{Halliday}, I., {Griffin}, A.~A., {Blackwell}, A.~T., Mar. 1996. {Detailed data
  for 259 fireballs from the Canadian camera network and inferences concerning
  the influx of large meteoroids}. Meteoritics and Planetary Science 31,
  185--217.

\bibitem[{{Hoppe}(1937)}]{Hoppe1937}
{Hoppe}, J., Apr. 1937. {Die physikalischen Vorg{\"a}nge beim Eindringen
  meteoritischer K{\"o}rper in die Erdatmosph{\"a}re}. Astronomische
  Nachrichten 262, 169--198.

\bibitem[{{Kero} et~al.(2012){Kero}, {Fujiwara}, {Abo}, {Szasz}, and
  {Nakamura}}]{Kero2012}
{Kero}, J., {Fujiwara}, Y., {Abo}, M., {Szasz}, C., {Nakamura}, T., Aug. 2012.
  {MU radar head echo observations of the 2011 October Draconids}. Monthly
  Notices of the Royal Astronomical Society 424, 1799--1806.

\bibitem[{{Levin}(1961)}]{Levin1961}
{Levin}, B.~I., 1961. {Physikalische Theorie der Meteore und die meteoritische
  Substanz im Sonnensystem.} Akademie-Verlag, Berlin.

\bibitem[{{Levin}(1956)}]{Levin1956}
{Levin}, B.~Y., 1956. {Fizicheskaia teoriia meteorov i meteorne veshchestvo v
  Solnechnoi sisteme (Physical Theory of Meteors and Meteorite Susbtance in the
  Solar System)}. Akad. Nauk SSSR, Moscow (in Russian).

\bibitem[{{McCrosky} and {Boeschenstein}(1965)}]{McCrosky1965}
{McCrosky}, R.~E., {Boeschenstein}, Jr., H., May 1965. {The Prairie Meteorite
  Network}. SAO Special Report 173.

\bibitem[{{McCrosky} and {Posen}(1968)}]{McCrosky1968}
{McCrosky}, R.~E., {Posen}, A., May 1968. {Special Data-Reduction Procedures
  for Prairie Network Meteor Photographs}. SAO Special Report 273.

\bibitem[{{Pecina} and {Ceplecha}(1983)}]{Pecina1983}
{Pecina}, P., {Ceplecha}, Z., Mar. 1983. {New aspects in single-body meteor
  physics}. Bulletin of the Astronomical Institutes of Czechoslovakia 34,
  102--121.

\bibitem[{{Popelenskaya}(2010)}]{Popelenskaya2010}
{Popelenskaya}, N.~V., 2010. {Dependence of the height of disappearance for
  small meteoric bodies on their parameters}. Moscow University Mechanics
  Bulletin 65~(4), 90--93.

\bibitem[{{Revelle}(1979)}]{Revelle1979}
{Revelle}, D.~O., May 1979. {A quasi-simple ablation model for large meteorite
  entry - Theory vs observations}. Journal of Atmospheric and Terrestrial
  Physics 41, 453--473.

\bibitem[{{Stulov}(1997)}]{Stulov1997}
{Stulov}, V.~P., Nov. 1997. {Interactions of space bodies with atmospheres of
  planets}. Applied Mechanics Reviews 50, 671--688.

\bibitem[{{Stulov}(1998)}]{Stulov1998}
{Stulov}, V.~P., Feb. 1998. {Gasdynamical model of the Tunguska fall}.
  Planetary and Space Science 46, 253--260.

\bibitem[{{Stulov}(2004)}]{Stulov2004}
{Stulov}, V.~P., 2004. Asymptotic theory of the motion of celestial bodies in
  the atmosphere. Planetary and Space Science 52~(5–6), 459 -- 463.

\bibitem[{{Stulov} et~al.(1995){Stulov}, {Mirskii}, and {Vilsyi}}]{Stulov1995}
{Stulov}, V.~P., {Mirskii}, V.~N., {Vilsyi}, A.~I., 1995. {Aerodinamika bolidov
  (Aerodynamics of Bolides)}. Nauka, Moscow (in Russian).

\bibitem[{{Vaubaillon} et~al.(2015){Vaubaillon}, { Koten}, {Margonis}, {Toth},
  {Rudawska}, {Gritsevich}, {Zender}, { McAuliffe}, {Pautet}, {Jenniskens},
  {Koschny}, {Colas}, {Bouley}, {Maquet}, {Leroy}, {Lecacheux}, {Borovicka},
  {Watanabe}, and {Oberst}}]{Vaubaillon2015}
{Vaubaillon}, J., { Koten}, P., {Margonis}, A., {Toth}, J., {Rudawska}, R.,
  {Gritsevich}, M., {Zender}, J., { McAuliffe}, J., {Pautet}, P.~D.,
  {Jenniskens}, P., {Koschny}, D., {Colas}, F., {Bouley}, S., {Maquet}, L.,
  {Leroy}, A., {Lecacheux}, J., {Borovicka}, J., {Watanabe}, J., {Oberst}, J.,
  2015. {The 2011 Draconids: the first European airborne meteor observation
  campaign}. Earth, Moon, and
  Planets.Http://dx.doi.org/10.1007/s11038-014-9455-5.

\bibitem[{{Wetherill} and {Revelle}(1981)}]{Wetherill1981}
{Wetherill}, G.~W., {Revelle}, D.~O., Nov. 1981. {Which fireballs are
  meteorites - A study of the Prairie Network photographic meteor data}. Icarus
  48, 308--328.

\bibitem[{{Whipple} and {Jacchie}(1957)}]{Whipple1957}
{Whipple}, F.~L., {Jacchie}, L.~G., 1957. {Reduction Methods for Photographic
  Meteor Trails}. Smithsonian Contributions to Astrophysics 1, 183--206.

\bibitem[{{Ye} et~al.(2013){Ye}, {Brown}, {Campbell-Brown}, and
  {Weryk}}]{Ye2013}
{Ye}, Q., {Brown}, P.~G., {Campbell-Brown}, M.~D., {Weryk}, R.~J., Nov. 2013.
  {Radar observations of the 2011 October Draconid outburst}. Monthly Notices
  of the Royal Astronomical Society 436, 675--689.

\end{thebibliography}




%%\bibitem[ ()]{}

%%\end{thebibliography}
\end{document}